\documentclass[conference, dvipsnames]{IEEEtran}
\usepackage{cite}
\usepackage{amsmath,amssymb,amsfonts}
\usepackage{algorithmic}
\usepackage{graphicx}
\usepackage{textcomp}
\usepackage{xcolor}
\usepackage{xcolor}
\usepackage{listings}
\usepackage{booktabs}
\usepackage{dirtytalk} 
\usepackage{makecell}
\usepackage{epstopdf}
\usepackage{graphics}
\usepackage{tikz}
\usepackage{graphicx}
\usepackage{rotating}
\usepackage{hyperref}
\usepackage{pgf-pie} 
\usepackage{listings}
\usepackage[final]{pdfpages}
\usepackage{media9}
\usepackage{lipsum}
\usepackage{pgfplots}
\usepackage{enumitem}
\usepackage{subcaption}
\usepackage{multirow}
\usepackage{multicol}
\usepackage{subcaption,siunitx,booktabs}
\usepackage{rotating}
\usepackage{array}
\usepackage{multirow}
\usepackage{tabularx}

\def\BibTeX{{\rm B\kern-.05em{\sc i\kern-.025em b}\kern-.08em
    T\kern-.1667em\lower.7ex\hbox{E}\kern-.125emX}}

\newcommand{\mpir}[0]{\textsc{MPI-rical}}
\newcommand{\mpicodecorpus}[0]{\emph{MPICodeCorpus}}

\DeclareRobustCommand{\IEEEauthorrefmark}[1]{\smash{\textsuperscript{\footnotesize #1}}}

\usepackage{todonotes}

\begin{document}

\title{{\mpir}: Data-Driven MPI Distributed Parallelism Assistance with Transformers}
\author{\IEEEauthorblockN{
Nadav Schneider\IEEEauthorrefmark{1,2},
Tal Kadosh\IEEEauthorrefmark{2,3},
Niranjan Hasabnis\IEEEauthorrefmark{4},
Timothy Mattson\IEEEauthorrefmark{4},
Yuval Pinter\IEEEauthorrefmark{3} and
Gal Oren\IEEEauthorrefmark{5,6}}\\
\IEEEauthorblockA{\IEEEauthorrefmark{1}Department of Electrical \& Computer Engineering, Ben-Gurion University, Israel}
\IEEEauthorblockA{\IEEEauthorrefmark{2}Israel Atomic Energy Commission}
\IEEEauthorblockA{\IEEEauthorrefmark{3}Department of Computer Science, Ben-Gurion University, Israel}
\IEEEauthorblockA{\IEEEauthorrefmark{4}Intel Labs, United States}
\IEEEauthorblockA{\IEEEauthorrefmark{5}Scientific Computing Center, Nuclear Research Center – Negev, Israel}
\IEEEauthorblockA{\IEEEauthorrefmark{6}Department of Computer Science, Technion – Israel Institute of Technology, Israel}
{\tt\small nadavsch@post.bgu.ac.il, talkad@post.bgu.ac.il, niranjan.hasabnis@intel.com,}\\ {\tt\small timothy.g.mattson@intel.com, pintery@bgu.ac.il, galoren@cs.technion.ac.il}
}


\maketitle

\begin{abstract}
Computational science has made rapid progress in recent years, leading to ever increasing demand for supercomputing resources. For scientific applications that leverage such resources, Message Passing Interface (MPI) plays a crucial role in enabling distributed memory parallelization across multiple nodes. However, parallelizing MPI code manually, and specifically, performing \emph{domain decomposition}, is a challenging and error-prone task.

In this paper, we address this problem by developing {\mpir}, a novel data-driven, programming-assistance tool that assists programmers in writing domain decomposition based distributed memory parallelization code using MPI. Specifically, we leverage Transformer architecture --- the invention that led to advancements in the field of natural language processing (NLP) --- with a supervised language model to suggest MPI functions and their proper locations in the code on the fly.
In addition to the novel model for MPI-based parallel programming, in this paper, we also introduce {\mpicodecorpus}, the first publicly-available corpus of MPI-based parallel programs that is created by mining more than 15,000 open-source repositories on GitHub.
Experimental results demonstrate the effectiveness of {\mpir} on both dataset from {\mpicodecorpus} and more importantly, on a compiled benchmark of MPI-based parallel programs for numerical computations that represent real-world scientific applications. Specifically, {\mpir} achieves F1 scores between 0.87-0.91 on these programs, demonstrating its accuracy in suggesting correct MPI functions at appropriate code locations. The source code used in this work, as well as other relevant sources, are available at: \textcolor{blue}{\url{https://github.com/Scientific-Computing-Lab-NRCN/MPI-rical}}.
\end{abstract}



\section{Introduction}
Computational science is a research field that simulates scientific problems with mathematical or analytical models, usually on supercomputers~\cite{prabhu2011survey}.
Some simulations are mostly phenomena too complex to be reliably predicted by theory analytically and too dangerous or expensive to reproduce in a lab. In order to imitate these complex phenomena properly, their corresponding simulations require high resolution, hence, a considerable amount of memory and CPU hours.
These massive calculations are typically parallelized via the distributed memory paradigm across multiple nodes---the overall calculation is solved by dividing it among different processors that communicate with each other and are individually responsible for a local yet synchronized sub-calculation.
The concept of communication and a shared computation between different processors is described by the Message-Passing model.
This enables the full exploitation of the given computing power by using all the available memory. Nowadays, the most common parallelization paradigm addressing these tasks is MPI with its popular implementations --- OpenMPI and MPICH~\cite{pacheco1997parallel}.

As mentioned, to simulate complex phenomena accurately via scientific applications, high resolution, which leads to high memory consumption, is required. Such complex simulations usually cannot be performed on a single node due to their insufficient memory and/or computational power. Consequently, scientific applications have been adopting the distributed computations approach. One of the common methods for distributed computing is to partition the data structures into different nodes, such that each node holds a different partition of the data structures. This method, known as \emph{domain decomposition}~\cite{smith1997domain},
solves the problem by splitting the problem's \emph{domain} (data structures, for example) into subdomains. Although each subdomain performs the same computations on different partitions of the data, these computations are performed in parallel to other subdomains, thus reducing the time required to perform complex simulations.

Successes in computational science have sharply increased the demand for supercomputing resources over the past twenty years~\cite{usingmpi} and, therefore, there is an ever-present need for distributed parallelization schemes. Consequently, many automated static tools for source-to-source parallelization of serial code to shared memory and shared memory to distributed memory have been created (Section~\ref{sec:previous_work}). Nonetheless, due to the complexity of the task at hand, none have been developed for translating serial code to distributed memory code. There are, inherently, more problems in parallelizing code in a distributed memory environment than in a shared-memory environment, specifically for distributing memory between different processes (Domain Decomposition). For instance, it is common to misplace send/receive functions, especially in a large source code or when a programmer is unfamiliar with the whole code. Moreover, one must know variable dependencies and the code structure from start to end; therefore, a deep understanding of the source code is required.


We believe that the emerging field of code language processing (CLP) shows the potential of building a code advisor for distributed memory parallelism. Code Language Processing is an emerging field that involves the application of probabilistic techniques such as AI and machine learning and deterministic techniques such as formal program analysis to understand code semantics, extract information from source code, and identify patterns, among others~\cite{a16010053, sharma2022survey, gottschlich2018three}. More importantly, the vast amount of open-source code hosted on code hosting sites such as GitHub, with access to powerful computing resources, has fueled the field of CLP. Recent advancements in the field of NLP have led to several tools that assist developers in various software engineering tasks such as analyzing comments, generating documentation from code, code summarization, code completion, and code translation~\cite{code_completion, shi2022natural, sym14030471}.
Some of these tools, such as GitHub's CoPilot~\footnote{\href{https://github.com/features/copilot}{https://github.com/features/copilot}}~\cite{githubcopilot}, and Google's code completion model~\cite{googleadvisor} also assist developers in their programming task by acting as code advisors (via plugins) to popular code editors (such as Visual Studio Code).

Over the years, different methods of translating OpenMP code into MPI code have been proposed, but none have been developed for translating serial code into parallel code for distributed memory systems. Nonetheless, there has been a recent attempt to address a related but different issue. Specifically, with the advancements in AI-based NLP, there have been attempts at generating MPI-based parallel code directly from natural language descriptions using large language models (LLMs)~\cite{chen2023lm4hpc, godoy2023evaluation, xu2022systematic}. LLMs, as the name suggests, are AI models that are trained to understand natural languages using large training datasets. These LLMs are then fine-tuned on source code-related training datasets (to produce code LLMs) and are evaluated on various downstream tasks such as code completion, code summarization, etc. Nichols et al.~\cite{nichols2023modeling} recently fine-tuned PolyCoder~\cite{xu2022systematic}, a code LLM, on a high-performance computing (HPC) dataset and evaluated it on the downstream tasks of automatically generating OpenMP-based and MPI-based programs from a given natural language prompt. Note, however, that none of the existing approaches attempt to translate serial code into MPI-based parallel code.

In this paper, we approach the problem of an MPI-based domain decomposition code assistance by using a code LLM. 
Specifically, we suggest appropriate MPI functions as well as their code locations to insert those functions to the programmer. Towards that end, we propose a novel approach that leverages a Transformer-based code LLM, named SPT-Code, that has been trained on semantic understanding tasks for several programming languages. Nevertheless, we apply SPT-Code for our particular task by finetuning it on a dataset of approximately 25,000 C programs out of more than 50,000 C programs from our {\mpicodecorpus}. We collected these programs from 16,500 GitHub repositories specifically for our task. We call the SPT-Code model finetuned on our dataset {\mpir}. Our experimental evaluation shows that {\mpir} performs well on our test dataset and, more importantly, on the benchmark of valid domain decomposition programs obtained from numerical computation codes (\autoref{compiled_bench}).
We believe {\mpir} can be deployed in code editors to assist programmers in the complicated and error-prone work of writing MPI-based parallel code for distributed memory systems.

{\textbf{Contributions.}} The main contributions of this paper are:
\begin{itemize}
    \item To the best of our knowledge, ours is the first approach that attempts to learn a data-driven programming assistance tool for distributed memory systems (and specifically MPI).
    \item We train and evaluate our approach, named {\mpir}, on 25,000 C programs obtained from GitHub and find that our model performs well in suggesting MPI functions for domain decomposition into an MPI-based parallel code. More importantly, we also evaluate {\mpir} on a set of numerical computations with domain decomposition programs and find that the model can also handle general programs well.
    \item Unlike existing approaches that directly generate MPI-based programs from natural language prompts, {\mpir} can be integrated into an IDE as an MPI domain decomposition assisting tool. It can thus serve as an in-editor programming assistance tool for MPI-based parallel programs.
    \item As a part of our research, we have created {\mpicodecorpus}, a large corpus of domain decomposition-based programs and their corresponding MPI-based parallel programs, obtained by mining 16,500 GitHub repositories. In the spirit of advancing scientific research, we are open-sourcing this corpus. To the best of our knowledge, {\mpicodecorpus} is the first publicly-available corpus of MPI-based parallel programs.
\end{itemize}

\section{Previous Work}
\label{sec:previous_work}

While we are unaware of any existing approaches for automatically translating serial code into parallel code for distributed memory systems, translating serial code into parallel code for shared memory systems has been an active area of research. 

\subsection{Serial to Shared-Memory Parallelization}
\label{sec:pre_serial_to_shared}
Techniques for the problem of serial code to shared-memory parallel code have mostly focused on heuristics and rule-based methods for many years. Examples include S2S compilers such as Cetus~\cite{dave2009cetus} and Par4All~\cite{creusillet2009par4all} that have been developed to automatically insert OpenMP \textit{pragma}s into code. Standard compilers such as GCC and LLVM also now contain passes for automatic parallelization of serial code. These methods often had limited capabilities and robustness. However, with the rapid growth of deep learning in general and specifically in the NLP domain, there have been several attempts~\cite{harel2023learning, kadosh2023advising} to apply data-driven language models to parallelize serial code for shared-memory systems automatically. More importantly, these probabilistic approaches improve upon deterministic tools such as compilers by handling incomplete code (such as code under development) and enabling the possibility of assisting programmers in code editors.

\subsection{Serial to Distributed-Memory Parallelization}
\label{sec:pre_serial_to_distributed}
Although, we are not aware of any existing approaches for translating serial code into MPI-based parallel code, there exists two types of techniques for generating MPI-based parallel code from different types of inputs.

\emph{(i) Generating MPI-based parallel code from OpenMP-based parallel code:}
Over the years, different methods of converting OpenMP-based parallel code into MPI-based parallel code have been proposed~\cite{basumallik2005towards, ferner2013paragin, garriga2015omp2mpi}, but none have been developed for translating serial code to MPI-based distributed memory code. There are, inherently, more problems in parallelizing code in a distributed memory environment than in a shared-memory environment, specifically for distributing memory between different processes (Domain Decomposition). It is common to misplace send/receive functions, especially in a large source code or when a programmer is unfamiliar with the whole code. One must know variable dependencies and the code structure from start to end; therefore, a deep understanding of the source code is required. The problem of translating OpenMP-based parallel code into MPI-based parallel code is relatively simpler since the question of if a structured block can be parallelized is already solved by OpenMP.

\emph{(ii) Generating MPI-based parallel code from natural language descriptions:} Recent advancements in deep-learning-based NLP such as Transformer architecture~\cite{vaswani2017attention} have led to the creation of large language models (LLMs), such as ChatGPT, that can now perform various natural language tasks such as question answering, translation, etc. Various existing approaches have finetuned these LLMs on source code datasets~\cite{xu2022systematic}, including even those specific to HPC to solve HPC related tasks such as predicting OpenMP pragmas for code~\cite{chen2023lm4hpc, godoy2023evaluation, nichols2023modeling}. Nichols et al.~\cite{nichols2023modeling} recently finetuned GPT-2, GPT-Neo, and PolyCoder LLMs~\cite{xu2022systematic} using their HPC dataset (source programs in C/C++ and repositories filtered by HPC related topics) to predict OpenMP pragmas for loops and to generate MPI-based parallel programs from natural language problem description. They named their best-performing model (i.e., PolyCoder trained on HPC source code) HPC-Coder. Although HPC-Coder shows promising results in generating MPI-based parallel programs, it is unclear if it can generate MPI-based programs for problem descriptions that are not part of its training dataset. This is because the test MPI programs that they used for their evaluation are well-known problems having existing MPI-based implementations. We believe that unseen problem descriptions would be the case of MPI programmers implementing domain decomposition in serial code in code editors.


\section{Research Objectives}
\label{research_objective}

Our objective in this research effort is to develop a programming-assistance tool that can assist MPI programmers in automatically generating correct MPI functions in an MPI-based domain decomposition parallel code. Developing this tool implies the model has an understanding of MPI routines. We break this problem down into two subproblems:


\subsection*{\normalsize{RQ1: \textit{\textbf{Is {\mpir} capable of generating calls to correct MPI functions?}}}}
Conceptually, the difference between the underwriting domain decomposition MPI code and MPI code after our model suggestions is the list of MPI functions that would be called. Out of the many MPI functions called through the code, checking whether the right MPI functions have been generated is critical for assessing the model's suggested solution. Note that in this version of the paper, we examine for functions' names only --- we do not consider function arguments.

\subsection*{\normalsize{RQ2: \textit{\textbf{Is {\mpir} capable of inserting the calls to MPI functions in the right locations?}}}}
Generating correct MPI functions is not enough because the locations of these functions in the code are also important for the validity of MPI programs. As such, this research question evaluates if the MPI function calls are inserted at the correct locations in the code. Conceptually, a location in code can be considered as a line number in a source program, an edge in the control-flow and a data-flow graph of a program, among others. In this effort, we keep it simple and consider the line number in the source program as the location.

\section{{\mpir}}
\label{sec:model}

\begin{figure*}[!ht]
 \centering
\begin{subfigure}{0.55\textwidth}
\centering
    \includegraphics[width=\textwidth]{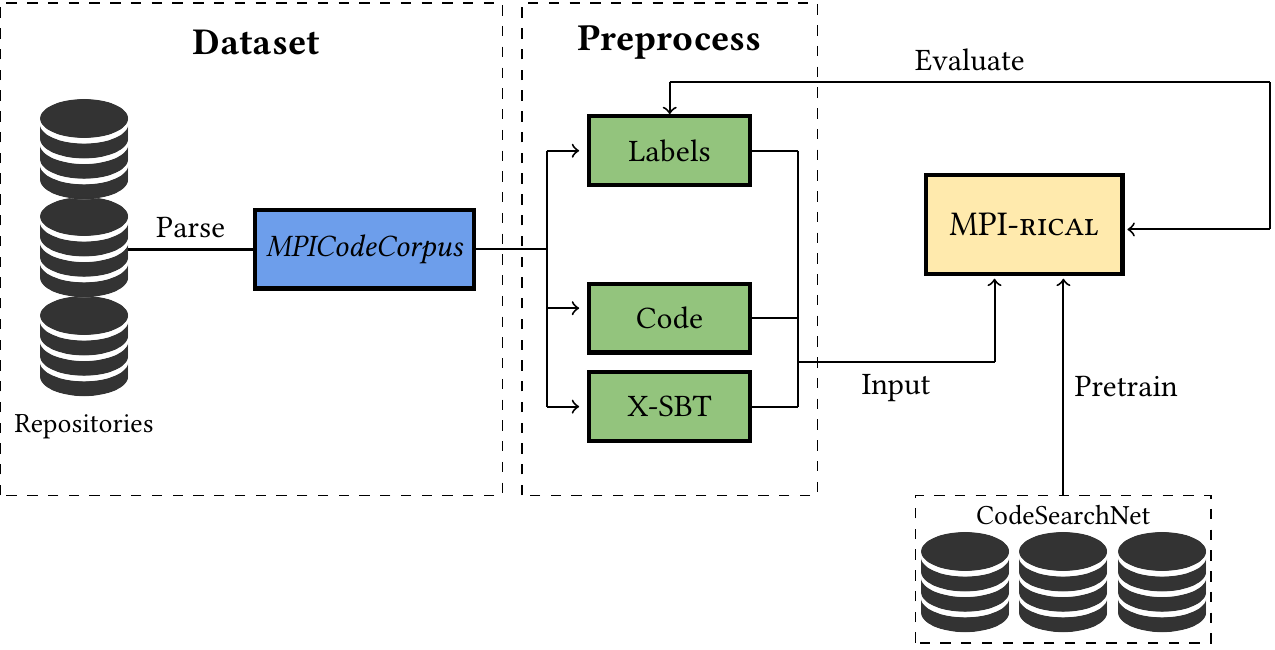}
    \caption{Overview of the model's training and evaluation. The dataset is created from {\mpicodecorpus} while three files constitute one example; MPI-based parallel C code (label), MPI-based parallel C code, with functions excluded, and its X-SBT (linearized AST). Our model, {\mpir}, trains and evaluates these examples. {\mpir} was pre-trained from the CodeSearchNet dataset.}
    \label{fig:mpir}
\end{subfigure}
\hfill
\begin{subfigure}{0.35\textwidth}
\centering
      \includegraphics[width=\textwidth]{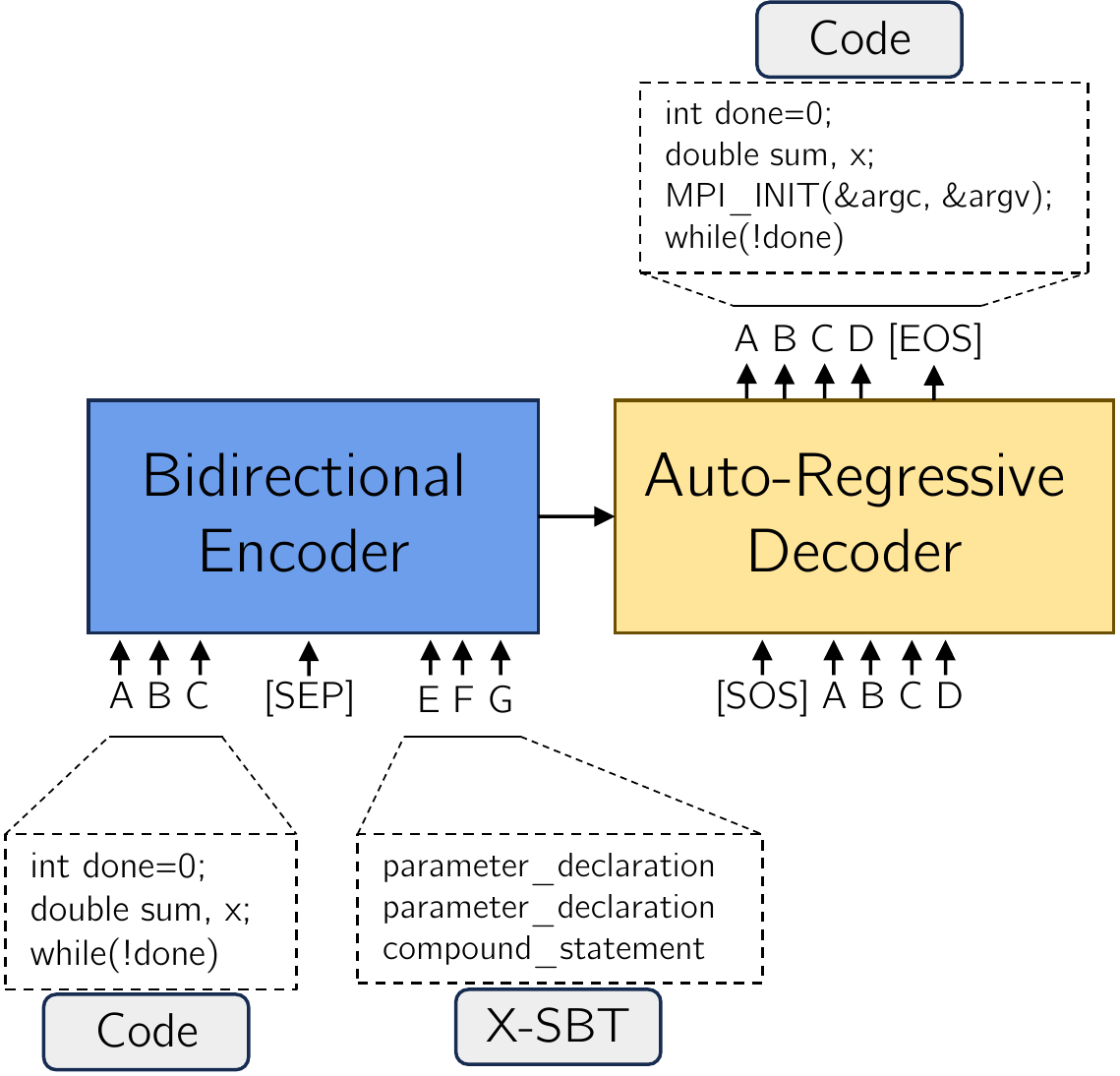}
    \caption{Translation task overview of {\mpir} (based on SPT-Code). The source code and its X-SBT are tokenized and concatenated into the encoder. Source code with the predicted MPI functions in the predicted locations is then derived out of the decoder --- The procedure is the same as the regular sequence-to-sequence model.}
    \label{fig:translation_task}
\end{subfigure}

\caption{Overview of {\mpir}}
\label{fig:overview}
\end{figure*}

Given the promising results delivered by Transformer architecture and, more specifically, SPT-Code~\cite{niu2022spt} model for the code translation problem, we chose SPT-Code for our approach. However, as SPT-Code does not support C/C++ languages or HPC programs, We fine-tune SPT-Code on a dataset created from {\mpicodecorpus}, a task-specific dataset that we have collected. This fine-tuned model, called {\mpir}, addresses the given task. This approach, as well as the chosen transformers-based model, will be explained in the following subsections.

\subsection{A brief background on SPT-Code}
SPT-Code is a multi-layer Transformer used by pre-training models such as BART~\cite{lewis2019bart}, and T5~\cite{raffel2020exploring} with source code in different languages in CodeSearchNet dataset~\cite{husain2019codesearchnet}, mainly Java and Python. SPT-Code has several innovations which led to achieving state-of-the-art performance on code-related downstream tasks after fine-tuning. Some of them are pre-training the decoder in addition to the encoder, which is often lacking with no real reason, and inputting the model with three different types of code components: plain code, linearized abstract-syntax tree (AST), and natural language, while linearized AST is natural language description of the AST. SPT-Code is significantly smaller than LLMs, enabling fast training/inference times, which is necessary for fusion in IDEs.

\textbf{\textit{Linearized AST.}}
AST, being a tree structure, cannot be fed directly to the SPT-Code model. As a result, AST needs to be converted into a linearized format. Towards that end, the SPT-Code authors develop a representation called X-SBT to represent the structural information of the source code. X-SBT is a simplified version of SBT~\cite{hu2018deep}. SPT-Code authors observe that sequences obtained by classical traversal methods (such as depth-first search) are lossy since AST can not be reconstructed back from them, meaning there are multiple labels to a certain input. This ambiguity can confuse the learning process. Therefore, SBT was created --- a structure-based traversal method to traverse the AST that addresses the specified problem. SPT-Code authors observed that SBT sequences, however, can be more than 3 times longer than the original code; hence, they developed X-SBT, which can reduce the length of the traversals' sequences by more than half. This is done by changing the sequence to an XML-like form and keeping syntactic information (expression level nodes and above) only. AST is parsed using \textit{TreeSitter} parser~\footnote{\textcolor{blue}{\url{https://github.com/tree-sitter/tree-sitter}}}. Note that \textit{TreeSitter} quickly and successfully parses code even if there are syntax errors. These advantages make \textit{TreeSitter} suited to live code advising.

\begin{sidewaysfigure*}[!htbp]
\centering
      \includegraphics[width=\textwidth]{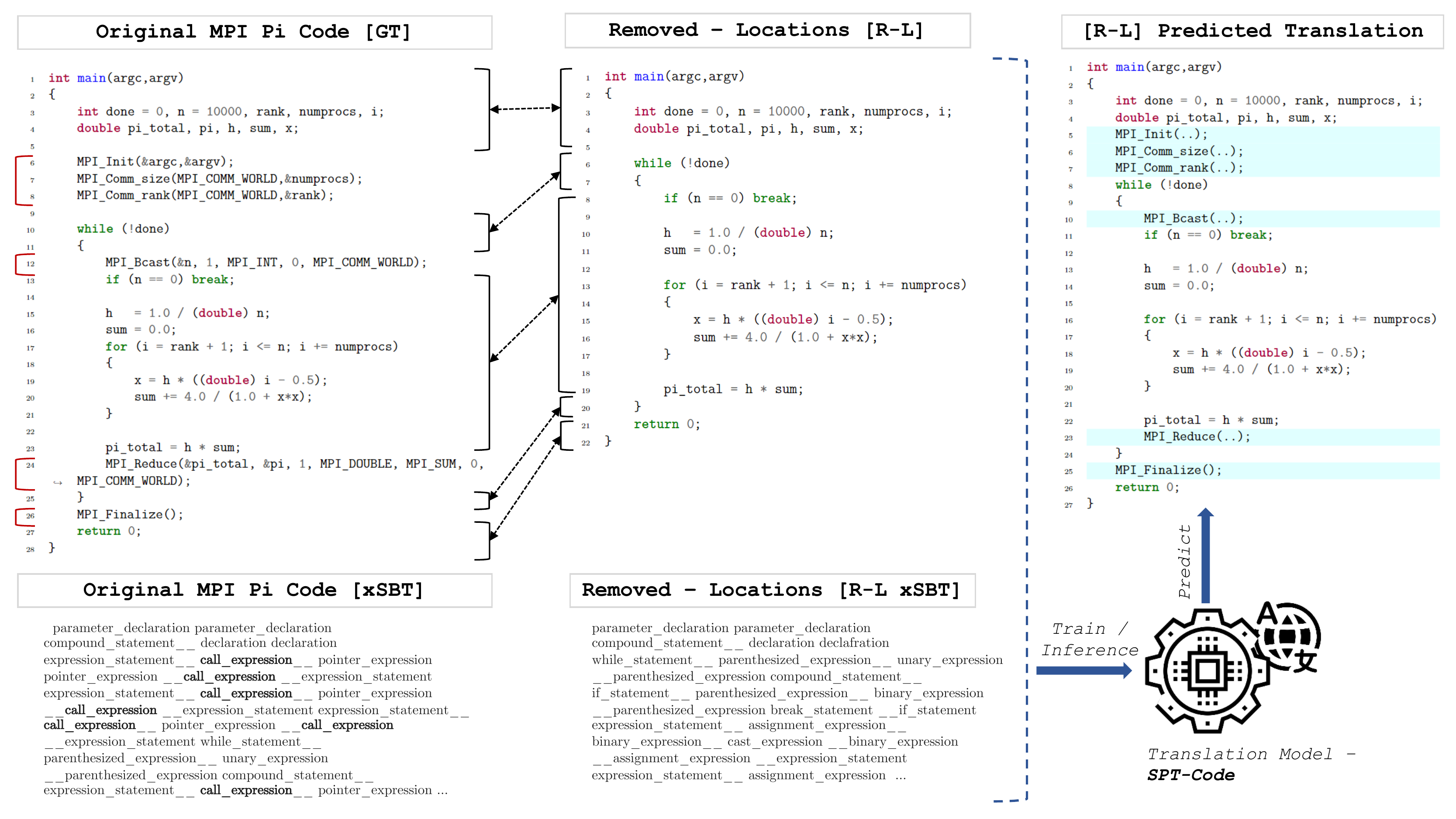}
    \caption{A data-oriented workflow of the \mpir{} system, exemplified by a distributed Pi calculation code: Given a corpus of MPI codes, a subset of the original dataset is created --- \textsc{Removed - Locations [R-L]} --- in which the MPI functions are replaced by a void and the original locations are not preserved. This is also done to the xSBT representations. Accordingly, the subset [R-L] is used to train a translation model, \textsc{SPT-Code}, which will eventually predict the desired MPI classification for new samples of codes given in those fashions. As such, \mpir{} is useful in easing the writing of MPI codes in IDEs.}
    \label{fig:mpichart}
\end{sidewaysfigure*}

\subsection{Fine-Tuning --- Code Translation}
\label{code_translation}
Conceptually, we consider our task as a code translation problem, translating MPI-based domain decomposition code without MPI functions to include MPI functions. Code translation handles both the research objectives, generating MPI functions and predicting their locations. Code translation in SPT-Code works as a generation task in a regular sequence-to-sequence fashion, while the input to the encoder is the source code, and then its linearized AST is separated by a symbol [SEP].

Figure~\ref{fig:overview} shows the overview of {\mpir}. Specifically, Figure~\ref{fig:mpir} shows the training and evaluation process of {\mpir}. In particular, we mine GitHub repositories for MPI-based C/C++ programs for HPC to build our dataset named {\mpicodecorpus}. We discuss the details of our dataset in the next section. Every C program from our dataset is then processed to produce input and output for {\mpir}. Specifically, as every C program from {\mpicodecorpus} contains MPI functions already, we consider them as \emph{labels}. In order to generate inputs for our model, we prune MPI function calls from input programs to generate MPI functions free programs (denoted as \emph{code} in the figure) and then generate their X-SBT using SPT-Code. Figure~\ref{fig:translation_task} shows the details of the inputs and outputs of our translation task. We illustrate the full workings of {\mpir} in \autoref{fig:mpichart}.


\subsection{Measuring performance of {\mpir} on the research objectives}

Although we formulate the given task as a code translation task, in order to measure the performance of {\mpir} we observe that both research objectives can be evaluated as classification tasks. We perform this simplification as both objectives can be nicely formulated as classification objectives. Specifically, the research question \textbf{RQ1} can be thought of as a multi-class classification problem --- predicting an MPI function out of 456 possible MPI functions (i.e., \emph{classes}) appearing in {\mpicodecorpus}. The second research question \textbf{RQ2} can be thought of as a binary classification problem --- given a code location (as a code token), predict if an MPI function would appear after it.

\section{{\mpicodecorpus}}

\subsection{Corpus}
\label{sec:corpus}

\begin{table}[!tbp]
    \begin{subtable}[t]{.20\textwidth}
    \centering
    \begin{tabular}[t]{lr}
    \toprule
    \textbf{\# Line}  & \textbf{Amount}  \\
    \midrule
     $\leq$ 10 & 2,670 \\ 
     11-50 & 22,361 \\
     51-99  & 14,078 \\
     $\geq$ 100 & 10,575 \\[1ex] 
    \bottomrule
    \end{tabular}
    \caption{Code lengths.}
    \label{table:line_length}
    \end{subtable}%
    \hfill
    \begin{subtable}[t]{.25\textwidth}
    \centering
    \begin{tabular}[t]{lr}
    \toprule
    \textbf{Function}  & \textbf{Amount}  \\
    \midrule
     \texttt{MPI\_Finalize} & 35,983 \\ 
     \texttt{MPI\_Comm\_rank} & 32,312 \\
     \texttt{MPI\_Comm\_size}  & 28,742 \\
     \texttt{MPI\_Init} & 25,114 \\
     \texttt{MPI\_Recv}  & 10,340 \\
     \texttt{MPI\_Send} & 9,841 \\
     \texttt{MPI\_Reduce} & 8,503 \\
     \texttt{MPI\_Bcast} & 5,296 \\
    \bottomrule
    \end{tabular}
    \caption{MPI Common Core functions (counted per file).}
    \label{table:functions_distribution}
    \end{subtable}%
        \hfill
\caption{General statistics related to the {\mpicodecorpus}.}
\end{table}

While several large code datasets in diverse programming languages exist~\cite{lu2021codexglue, lachaux2020unsupervised, hasabnis2021controlflag, yao2018staqc, markovtsev2018public}, datasets containing MPI-based parallel programs are rare. Therefore, we have created a corpus consisting of MPI-based parallel programs to train and evaluate SPT-Code on the given task. Specifically, we created the corpus by mining eligible repositories on \textit{github.com}, the popular code hosting service for open-source software development. In particular, we used \textit{github-clone-all}~\footnote{\href{https://github.com/rhysd/github-clone-all}{https://github.com/rhysd/github-clone-all}}, a script for mining repositories stored on GitHub. We have extracted C files from repositories containing the phrase ``MPI'' in the title, description, and the \textit{README} file. Overall, we extracted 59,446 C programs from approximately 16,500 repositories --- we define a program as a source file containing the \texttt{main} function, its headers, and implementations' files.

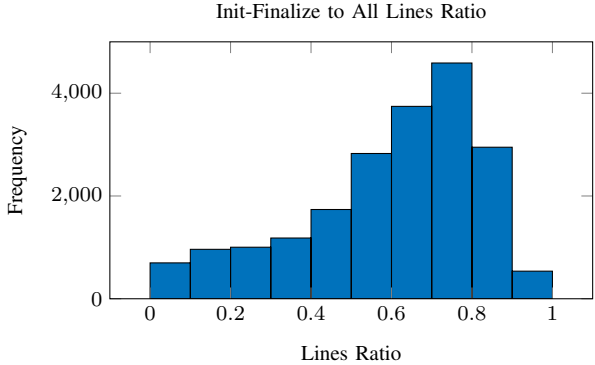
\begin{figure}[!tbp]
    \begin{footnotesize}
    \begin{tikzpicture}
    \begin{axis}[
        title=Init-Finalize to All Lines Ratio,
        xlabel=Lines Ratio,
        ylabel=Frequency,
        ymin=0, ymax=5000,
        area style,
        width=8cm,
        height=5cm
        ]
    \addplot+[black, ybar interval,mark=no, style={fill=RoyalBlue}] plot coordinates { (0.0, 698) (0.1, 962) (0.2, 1003) (0.3, 1182) (0.4, 1737) (0.5, 2827) (0.6, 3744) (0.7, 4588) (0.8, 2949) (0.9, 537) (1, 7)};
    \end{axis}
    \end{tikzpicture}
    \end{footnotesize}
    \caption{Ratio of the length of a parallel code to the overall program length}
    \label{fig:init_to_finalize_ratio}
\end{figure}

Moreover, we apply the following inclusion and exclusion criteria for dataset creation:

\subsubsection{Inclusion Criteria}
\label{sec:inclusion_criteria}
Once we extracted programs from the repositories, we parsed them using \textit{pycparser}~\cite{pycparser}, a Python package for generating an AST from C programs. In order to ensure that our model trains on complete MPI programs that contain all relevant MPI function calls, we only consider complete programs that are successfully parsed by the parser.

\subsubsection{Exclusion Criteria}
\label{sec:exclusion_criteria}
Due to hardware limitations, we have limited training examples to files with 320 tokens (approximately 50 lines). Hence, we excluded files with more than 320 tokens. In addition, the AST generation has been used as a selector. Programs that could not be parsed by \textit{pycparser} were excluded. Table~\ref{table:line_length} shows the distribution of code lengths of our corpus. Given the hardware limitation, we had to drop almost 50\% of the code examples from our corpus.

\subsubsection{Code Standardization}
Code standardization is crucial for both training the model and evaluating the results. We performed standardization by regenerating all the programs back from AST back --- amending wrong indentations, and deleting unnecessary linebreaks and spaces. 

\subsection{Dataset Creation}
\begin{figure*}[!tbp]
\centering
    \fontsize{6}{8} \selectfont 
        \scalebox{0.98}{\includegraphics{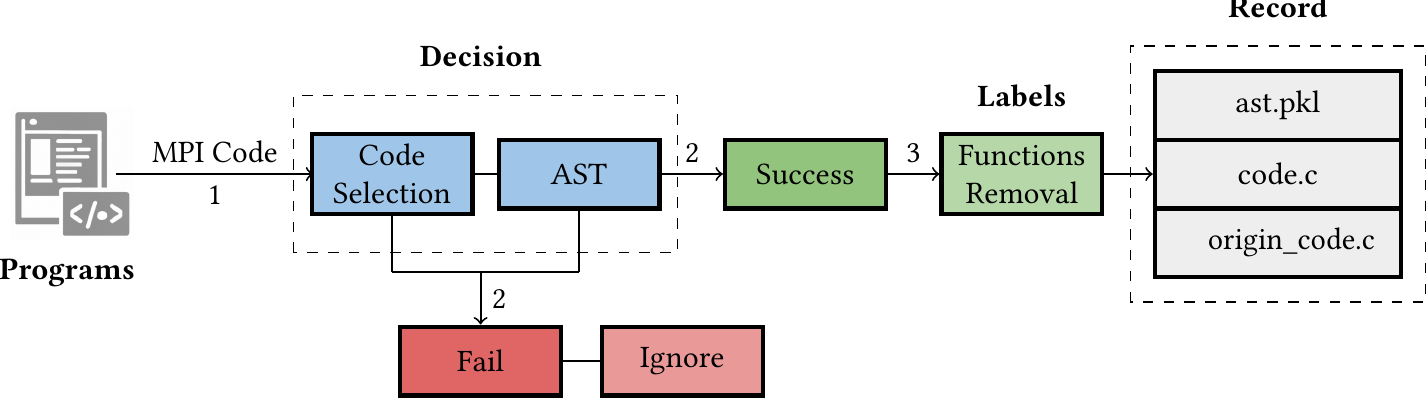}}
    \caption{Overview of the workflow for creating the dataset out of {\mpicodecorpus}; (1) programs from the corpus are extracted. (2) These programs pass through a selection stage. (3) MPI function calls from successful code snippets are removed. The resulting code snippets are then added to the dataset.}
    \vspace{-3ex}
    \label{fig:database_workflow}
\end{figure*}

Note that, as we mentioned before, although we have described our task as a code translation task, we measure the performance of {\mpir} as a classification task. Toward that end, we use our dataset for supervised learning. Our dataset contains MPI-based parallel code with MPI functions excluded as inputs while its corresponding original MPI-based parallel code as labels. Specifically, each MPI function in the MPI-based parallel code is replaced with an empty string (removed); hence, information about both functions and locations is lost. The overall process of creating our dataset is depicted in Figure~\ref{fig:database_workflow}. As mentioned before, the dataset contains approximately 25,000 examples.

\subsection{Statistics}

Understanding the most used functions, which we will call ``MPI Common Core'' functions, is critical for evaluating the model's performance. Ensuring that the model performs well on the relevant and most useful functions is a priority for validating that the model does its job and comprehends the given task. Towards that end, we analyzed {\mpicodecorpus} for MPI Common Core functions. The distribution of these functions is presented in~\autoref{table:functions_distribution}. Note that in the table, a function occurrence is counted per file, meaning multiple occurrences of some function in a file are still counted as one. One observation from the table is that the MPI function distribution is exponentially decreasing, while the MPI Common Core functions are at the beginning, while the rest are marginal.

Analyzing the ratio between the length of MPI-based parallel code and the whole code is important to understand the level of impact parallelization may have on the programs. In other words, we want to ensure that the parallelization consumes a reasonable amount of code and is not a minor addition. We determine the length of MPI-based parallel code by considering \texttt{MPI\_Init}, which initializes the parallelized area, and \texttt{MPI\_Finalize}, which ends it. Therefore, counting the number of lines between these two functions gives the length of parallel code. As can be seen from (\autoref{fig:init_to_finalize_ratio}), most of the MPI programs have more than half of the lines inside the parallelized code, which is sufficient for training. In addition, it is important to note that out of the raw data, \texttt{MPI\_Init} and \texttt{MPI\_Finalize} were both included in 20,228 files.

\section{Experimental Results}
\begin{figure}[!hb]
    \centering
    \hspace{-50pt}\begin{subfigure}{0.45\linewidth}
        \centering
        \begin{tikzpicture}
        \begin{axis}[
            grid style={line width=.1pt, draw=gray!10},
            major grid style={line width=.2pt,draw=gray!10},
            minor tick num=5,
            grid=both,
            title=\textbf{Training Loss},
            title style={yshift=-2.1ex,},
            legend style={nodes={scale=0.5, transform shape}},
            xlabel=\#Epochs,
            width=\linewidth
        ]
        \addplot [mark=triangle*, line width=1pt, red] table [y=train, x=epoch]{results/translation_placeholder.dat};
        \end{axis}
        \end{tikzpicture}
        \captionsetup{width=0.9\linewidth}
    \end{subfigure}\hspace{-25pt}
    \begin{subfigure}{0.45\linewidth}
        \centering
        \begin{tikzpicture}
        \begin{axis}[
            grid style={line width=.1pt, draw=gray!10},
            major grid style={line width=.2pt,draw=gray!10},
            minor tick num=5,
            grid=both,
            title=\textbf{Validation Loss},
            title style={yshift=-2.1ex,},
            legend style={nodes={scale=0.5, transform shape}},
            xlabel=\#Epochs,
            width=\linewidth
        ]
        \addplot [mark=triangle*, line width=1pt, red] table [y=valid, x=epoch]{results/translation_placeholder.dat};
        \end{axis}
        \captionsetup{width=0.9\linewidth}
        \label{fig:validloss}
        \end{tikzpicture}
    \end{subfigure}\hspace{-25pt}
    \begin{subfigure}{0.45\linewidth}
        \centering
        \begin{tikzpicture}
        \begin{axis}[
            grid style={line width=.1pt, draw=gray!10},
            major grid style={line width=.2pt,draw=gray!10},
            minor tick num=5,
            grid=both,
            xlabel=\# Epochs,
            title=\textbf{Accuracy},
            title style={yshift=-2.1ex,},
            legend style={nodes={scale=0.5, transform shape}},
            width=\linewidth
        ]
        \addplot [mark=triangle*, line width=1pt, red] table [y=accuracy, x=epoch]{results/translation_placeholder.dat};
        \end{axis}
        \end{tikzpicture}
        \captionsetup{width=0.9\linewidth}
        \label{fig:accuracy}
    \end{subfigure}\hspace{-30pt}
    
    \caption{The training loss, validation loss, and accuracy of {\mpir} as a function of epoch numbers.}
    \label{fig:trainvalidacc}
\end{figure}
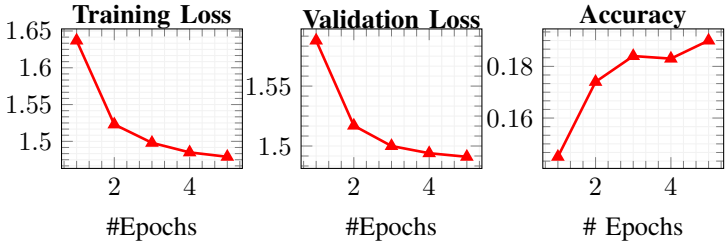

\textbf{\textit{Setup.}} We trained {\mpir} on our dataset containing 24,125 examples. We split this dataset into train, validation, and test using a standard splitting ratio of 80:10:10. We then train the model with the translation task. The model was developed using PyTorch~\cite{paszke2019pytorch} framework and was trained on Nvidia Tesla-V100 GPU, having 32GB memory. The training was carried out with a batch size of 32, 320 tokens and 5 epochs for each task. \autoref{fig:trainvalidacc} shows the progress of training runs over multiple epochs.

\subsection{Performance Metrics}
As we measure the performance of {\mpir} as a supervised classification task, we use standard metrics of F1 score, precision, and recall to measure the performance of {\mpir}.


We define positive and negative prediction for our task as:
\begin{enumerate}
    \item True Positive (TP) --- {\mpir} predicts that an MPI function can be inserted at a given location, and the predicted MPI function is indeed the ground truth (\emph{label}).
    \item False Positive (FP) --- {\mpir} predicts that an MPI function can be inserted at a given location, but the predicted MPI function is different than the ground truth (\emph{label}).
    \item True Negative (TN) --- The model predicts that no MPI function can be inserted at a given location, and this prediction matches with the ground truth.
    \item False Negative (FN) --- The model predicts that no MPI function can be inserted at a given location, but the ground truth says otherwise.
\end{enumerate}

\begin{figure*}
\centering
    \fontsize{6}{8} \selectfont 
        \scalebox{0.4}{\includegraphics{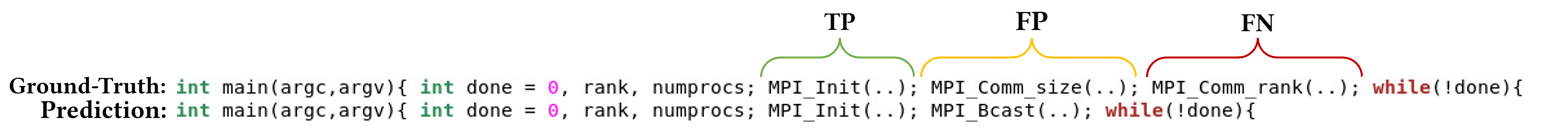}}
    \caption{Positive and negative prediction for measuring our models' performance is illustrated.}
    \vspace{2ex}
    \label{fig:f1_calc}
\end{figure*}

Our measurements have one-line tolerance, meaning identical ground-truth MPI function and its corresponding generated function will be considered matching only if there is one line difference between their locations. This is important since replacing the locations of two near MPI functions usually has no influence on the code. \autoref{fig:f1_calc} demonstrates the notion of positive and negative outcomes for our task. Note that TN is out of our scope since our focus is on MPI function generation.

\subsection{Results on \mpicodecorpus}

\begin{table}[tbp]
\centering
    \begin{tabular}{c|c} 
     \hline
     \textbf{Quality Measure}  & \textbf{{\mpicodecorpus}} \\ [0.5ex] 
     \hline \hline
     M-F1 & 0.87 \\ 
     \hline
     M-Precision & 0.85 \\ 
     \hline
     M-Recall & 0.89 \\ 
     \hline
     MCC-F1 & 0.89 \\ 
     \hline
     MCC-Precision & 0.91 \\ 
     \hline
     MCC-Recall & 0.87 \\ 
     \hline
     BLEU & 0.93 \\ 
     \hline
      Meteor & 0.62 \\ 
     \hline
     Rouge-l & 0.95 \\ 
     \hline
     ACC & 0.57 \\ 
     \hline
    \end{tabular}
    \caption{Performance of {\mpir} on {\mpicodecorpus} test set.}
    \label{table:results_mpicodecorpus}
\end{table}

Performance of {\mpir} on {\mpicodecorpus} is in~\autoref{table:results_mpicodecorpus}. MPI function distribution looks like a decreasing exponent. Therefore, it is important to distinguish between the commonly-used functions (MPI Common Core) and the rest. Hence, results have been evaluated on both the overall MPI functions and the MPI Common Core functions, marked as M-F1 and MCC-F1 correspondingly.

\subsection{Results on Fully Compiled Numerical Computations}
\label{compiled_bench}
After evaluating on {\mpicodecorpus}, we decided to evaluate {\mpir} on real-world benchmarks. Parallel programs in standard benchmarks are usually optimized and carefully tested. Therefore, they can offer insights into the model's performance on real, error-free MPI codes.

There are MPI-based parallel code examples from benchmarks such as Standard Performance Evaluation Corporation (SPEC)~\cite{dixit1991spec, juckeland2015spec}, NAS Parallel Benchmarks (NPB)~\cite{bailey1991parallel}, and even books like Using MPI~\cite{gropp1999using} and Using Advanced MPI~\cite{gropp2014using}. Nonetheless, only a small fraction of the code examples from these benchmarks passed our inclusion-exclusion criteria. Therefore, we compiled a benchmark of our own --- \textit{fully compiled numerical computation codes}~\footnote{\href{https://github.com/Scientific-Computing-Lab-NRCN/MPI-rical/tree/main/BENCHMARK}{https://github.com/Scientific-Computing-Lab-NRCN/MPI-rical/tree/main/BENCHMARK}} that are short and include complete code context (the examples and the results are attached to the footnote).
Specifically, we have written and compiled these 11 MPI-based parallel code examples with domain decomposition. Each demonstrates a selected numerical computation, such as pi calculation with Riemann Sum, matrix multiplication, function integration and etc. All these examples have passed our inclusion criteria. We evaluated the validity of MPI-based parallel programs generated by {\mpir} by compiling and running them.

Performance of {\mpir} on this benchmark as measured by the suited M-F1 score is presented in \autoref{table:results_math_problems}. The automatic M-F1 calculation is limited since it only has one-line tolerance, meaning generated MPI functions and the ground-truth function will be considered matching only if there is one line difference between their locations. In this case, manually measuring enables more accurate results. In summary, the model achieved M-F1 of 0.91, M-Precision of 0.98, and M-Recall of 0.86.



\begin{table}[!tb]
\centering
    \begin{tabular}{c|c|c|c} 
     \hline
      \textbf{Code} & \textbf{M-F1 }& \textbf{M-Precision} &\textbf{ M-Recall} \\ [0.5ex] 
     \hline \hline
     Array Average & 0.88 & 1.0 & 0.8 \\ 
     \hline
     Vector Dot Product & 0.88 & 1.0 & 0.8 \\ 
     \hline
     Min-Max & 0.66 & 1 & 0.5 \\ 
     \hline
     Matrix-Vector Multiplication & 0.9 & 0.83 & 1.0 \\ 
     \hline
     Sum (Reduce \& Gather) & 0.8 & 1.0 & 0.6 \\ 
     \hline
     Merge Sort & 1.0 & 1.0 & 1.0 \\ 
     \hline
     Pi Monte-Carlo & 1.0 & 1.0 & 1.0 \\ 
     \hline
      Pi Riemann Sum & 1.0 & 1.0 & 1.0 \\ 
     \hline
     Factorial & 0.88 & 1.0 & 0.8 \\ 
     \hline
     Fibonacci & 1.0 & 1.0 & 1.0 \\ 
     \hline
     Trapezoidal Rule (Integration) & 1.0 & 1.0 & 1.0 \\ 
     \hline
     \hline
     \textbf{Total} & \textbf{0.91} & \textbf{0.98} & \textbf{0.86} \\
     \hline
    \end{tabular}
    \caption{Performance of {\mpir} on numerical computations benchmark.}
    \label{table:results_math_problems}
\end{table}

\section{Conclusion and Future Work}
Recent advancements in language models have transformed static automatic parallelization tools into contextualized learning models. In this paper, we propose a novel approach for assisting MPI programmers in writing MPI-based parallel code for distributed memory systems, specifically domain decomposition. Our proposed approach suggests MPI functions in the right locations using a transformer-based model named {\mpir}, which is trained on the first MPI-specific code corpus called {\mpicodecorpus}. Experimental results demonstrate that {\mpir} performs well on test programs from {\mpicodecorpus}. More importantly, we evaluated {\mpir} on 11 carefully selected and compiled MPI programs with domain decomposition codes to find that {\mpir} also performs well on real-world programs. We believe that the results imply good domain decomposition and MPI-based parallelization understanding of {\mpir}.

While we present the initial approach and results for a data-driven distributed-memory parallelism advisor model, we belive that further research directions have been unlocked.
For future work, we plan to create a model trained and evaluated on complete codes with no length restriction. To do that while still training a transformer-based model, there is a need to either dramatically upgrade the hardware or use continual learning~\cite{biesialska2020continual}. This will enable full exploitation of {\mpicodecorpus}. In addition, we belive that pre-train a language model with a corpus of C programs instead of Java or Python might improve the results. Furthermore, the current work measures generated MPI functions while ignoring structural code changes, which we believe are an integral part of MPI-based parallel programs. Additional research about measuring the accuracy of generated MPI functions' arguments and MPI code structure has to be done.

\section*{Acknowledgment}
This research was supported by the Israeli Council for Higher Education (CHE) via the Data Science Research Center, Ben-Gurion University of the Negev, Israel; Intel Corporation (oneAPI CoE program); and the Lynn and William Frankel Center for Computer Science. Computational support was provided by the NegevHPC project~\cite{negevhpc} and Intel Developer Cloud~\cite{intel-cloud}. The authors thank Re'em Harel, Israel Hen, and Gabi Dadush for their help and support.

\bibliographystyle{IEEEtran}
\bibliography{IEEEabrv,sample-base.bib}

\end{document}